# Tunable Infrared Phonon Anomalies in Trilayer Graphene

Chun Hung Lui[1], Emmanuele Cappelluti[2,3], Zhiqiang Li[1], Tony F. Heinz[1*]

[1]Departments of Physics and Electrical Engineering, Columbia University, 538 West 120th Street, New York, NY 10027, USA
[2]Istituto dei Sistemi Complessi (ISC), CNR, U.O.S. Sapienza, v. dei Taurini 19, 00185 Rome, Italy
[3]Instituto de Ciencia de Materiales de Madrid (ICMM), CSIC, Cantoblanco, Madrid, Spain

*Corresponding author (email: tony.heinz@columbia.edu)

Trilayer graphene in both ABA (Bernal) and ABC (rhombohedral) stacking sequences is shown to exhibit intense infrared absorption from in-plane optical phonons. The feature, lying at ~1580 cm$^{-1}$, changes strongly with electrostatic gating. For ABC-stacked graphene trilayers, we observed a large enhancement in phonon absorption amplitude, as well as softening of the phonon mode, as the Fermi level is tuned away from charge neutrality. A similar, but substantially weaker effect is seen in samples with the more common ABA stacking order. The strong infrared response of the optical phonons and the pronounced variation with electrostatic gating and stacking-order reflect the interactions of the phonons and electronic excitations in the two systems. The key experimental findings can be reproduced within a simplified charged-phonon model that considers the influence of charging through Pauli blocking of the electronic transitions.

Trilayer graphene has attracted much recent attention because of its unusual electronic properties. This interest is further heightened by the existence of two stable allotropes, the ABA and ABC stacked trilayers (Fig. 1) [1-4], which exhibit very different characteristics from one another, as well as from monolayer and bilayer graphene. Indeed, the two trilayer allotropes have been shown to possess distinctive electronic and transport properties [1-19]. With respect to phonons, one would, however, expect the in-plane phonons in trilayer graphene to exhibit little sensitivity to stacking order (or changed layer thickness) because of the weak interlayer lattice coupling [20].

In contrast to such an expectation of weak structure dependence for phonons, we show here that the infrared (IR) spectra of in-plane optical phonons in trilayer graphene are dramatically modified both by stacking order and charge density. We observe noticeable IR absorption by the zone-center in-plane optical phonons in both trilayer allotropes around 1580 cm$^{-1}$, but with distinct characteristics and variation with electrostatic gating. As we tune the Fermi level to either the electron or hole side, we record a strong enhancement in the IR absorption amplitude accompanied by a significant softening (~15 cm$^{-1}$) of the phonon mode in ABC trilayers. In contrast, the ABA trilayers show much weaker phonon features and a less pronounced response to doping. We are able to reproduce the key experimental observations within a model based on charged-phonon theory. Our analysis shows that the observed IR response of the optical phonons arises from their interactions with the interband electronic transitions. The increased response and greater sensitivity to doping in the ABC graphene trilayers reflects the presence of stronger low-energy electronic transitions compared with the ABA system. Our exploration of the infrared phonon anomaly with respect to both the stacking and doping degrees of freedom reveals the significant interplay between the two variables in the system. Our study complements and extends the previous studies IR spectroscopy of bilayer [21, 22] and few-layer graphene [23] that were limited to probing the influence of only one of these characteristics.

In our experiment, graphene trilayer samples were prepared by mechanical exfoliation of kish graphite (Covalent Materials Corp.) on SiO$_2$(300nm)/Si substrates. The sample thickness and the stacking order were characterized by means of IR and Raman spectroscopy [1-4, 11]. All IR measurements in our experiment were performed using the National Synchrotron Light Source at Brookhaven National Laboratory (U12IR beam line). We further used the signature of the stacking order in the 2D Raman feature to map out the stacking domains of the trilayer samples by Raman imaging [2]. We chose those samples with large (>200 μm$^2$) homogeneous domains of ABA or ABC stacking for device fabrication. We made use of the polymer electrolyte (poly(ethylene oxide):LiClO$_4$) top gate, which allowed us to induce high doping densities (~10$^{13}$ cm$^{-2}$) in the trilayer graphene samples. Details of the device fabrication are provided in [3, 24].

We measured the IR transmission spectrum of the gated trilayers at near-normal incidence by comparing spectra taken from regions of the substrate with and without the trilayer graphene sample. Figs. 2(a)-(b) display the fractional change in transmission for ABA and ABC trilayers for different gate voltages ($V_g$). In the transmission spectra of both trilayers, we identify a sharp feature at ~1580 cm$^{-1}$, the energy of the zone-center optical phonons. The slope in the spectra arises from the trilayer electronic response, as well as from residual absorption of the polymer top gate for which lateral thickness variation makes the normalization process imperfect. The electronic contribution, arising from absorption by the direct electronic transitions in the graphene trilayer, has been discussed in detail in [2, 3, 25]. In order to focus on the phonon feature, we subtracted a smooth baseline using a polynomial fitting procedure.

In our experimental condition, *i.e.* a thin film of graphene on a SiO$_2$(300-nm)/Si substrate, the IR transmission is related to both the real and imaginary parts of the optical conductivity of trilayer graphene. We have applied a Kramers-Kronig constrained variational spectral analysis with the RefFit software developed by A. B. Kuzmenko [22, 26] to extract the real part of the optical conductivity $\Delta\sigma(\hbar\omega)$ of the trilayer samples, taking into account the known optical constants and thicknesses of the SiO$_2$ and Si layers. In the calculation, we neglect for simplicity the effect of the thin polymer top-gate layer in our devices.

Figs. 3(a)-(b) display the differential conductivity spectra $\Delta\sigma(\hbar\omega)$ in the range of 1430-1730 cm$^{-1}$ extracted from the corresponding baseline-corrected transmission spectra in Figs. 2(a)-(b). A striking feature of our results is the strong gate induced enhancement of the phonon absorption peak for the ABC trilayer graphene [Fig. 3(b)]. The ABC trilayer does not show any noticeable absorption peak at the charge neutrality (*CN*) point of $V_g = V_{CN}$. As $V_g$ is tuned away from $V_{CN}$, however, a clear absorption feature appears at the zone-center optical phonon energy (~1580 cm$^{-1}$) and grows dramatically with the gate bias. For the highest doping levels, the strength of the phonon absorption is comparable to that of the main electronic transition peak at 370 meV. In addition, we observe a red shift of the absorption peak by more than 10 cm$^{-1}$ for both electron and hole doping. We also observe the presence of a phonon absorption peak in ABA graphene trilayers [Fig. 3(a)]. The amplitude and frequency of the phonon feature in ABA trilayer, however, varies relatively weakly with $V_g$. Although both ABA and ABC trilayers have similar absorption strength at low doping, the ABC phonon absorption is more than five times greater than the ABA absorption at high doping. We have fit each $\Delta\sigma(\hbar\omega)$ spectrum in Figs. 3(a)-(b) with a single Lorentzian function and extract the phonon spectral weight and frequency as the integrated area and peak position of the fit function, respectively. (In contrast to measurements of the optical phonons in bilayer and other few-layer graphene systems [21-23], only slight asymmetry was observed in the spectral lineshapes and we do not need to introduce a Fano lineshape to describe the results.) The contrast in behavior of the two types of trilayers can be seen clearly in Figs. 4(a)-(b), which display the fit parameters as a function of gate voltage. We note that we did not observe any systematic dependence of the spectral linewidth on the gate bias for both types of trilayers.

The observation of strong phonon absorption in graphene trilayers, one that is of similar strength to the direct electronic transitions, is striking. Since graphene trilayers are non-polar materials, the electrons are evenly distributed among all the carbon atoms and do not exhibit any significant static dipole moment. Upon doping the additional charges are somewhat unequally distributed on the atomic sites. We estimate, however, that the resulting dipole moments are still two orders of magnitude smaller than what is needed to explain the observed strength of phonon absorption. For a proper explanation of the phenomenon, we must consider the interactions between the phonons and electrons in the system. These interactions also allow us to understand the correlation between enhancement of the electronic absorption and the phonon absorption, as well as the large red shift of the phonon absorption peak, with doping. When the IR active phonons are coupled to the electronic transitions, their frequency will be modified and some of the oscillation strength of the electronic transitions can be transferred to the phonon system. The phonons can be considered to be dressed by the electrons, which bestow additional oscillator strength on them. This so-called charged-phonon absorption [27-29] has also been observed in other systems, including the bilayer and few-layer graphene [21-23], C$_{60}$ compounds [30], graphite [31] and some organic materials [32].

For quantitative analysis of our experimental data, we applied charged-phonon theory [27-29] to calculate the spectral properties of the IR active phonons in the ABA and ABC trilayer systems [33]. This theory, applied successfully in bilayer graphene [28, 29], allows us to obtain the absorption strength and the frequency shift induced by the electron-phonon interaction for each phonon mode in trilayer graphene from the real parts of the current-phonon response function and the phonon self-energy, respectively. To capture the key features of the phonon spectra, we adopt a simple model of the electronic structure based on a tight-binding scheme of the electronic structure that includes only the dominant intralayer ($\gamma_0$) and interlayer ($\gamma_1$) coupling [Figs. 1(c)-(d)]. We assume that the only effect of the gating is to modify the electron (hole) population within a rigid-band scheme, neglecting the influence of any gate-induced electric field across the graphene layers. We shall see, despite the simplicity of this model it is able to account for all the main spectral features in our experiment.

In the absence of perpendicular electric fields, we can identify two IR active modes $E'_a$ and $E'_b$ for the ABA trilayer and a single IR active phonon mode $E'_u$ for the ABC trilayer [34]. Figs. 1(a)-(b) display the atomic displacements of $E'_b$ mode for ABA structure and $E'_u$ mode for ABC structure. In our analysis, we compare the properties of these two modes. The $E'_a$ mode in ABA trilayer, which has the same atomic displacement with $E'_b$ except that the atoms in the middle layer move in opposite direction, has much lower IR activity than $E'_b$ mode and can be disregarded in our analysis.

The calculated phonon spectral weight and frequency of the $E'_b$ mode in ABA structure and of $E_u$ mode in ABC structure are shown in Figs. 4(a)-(b). To fit the experimental data as a function of gate voltage, we assumed for both devices a top-gate capacitance of $C = 2.6$ μFcm$^{-2}$, yielding a charge density of $n \sim 2 \times 10^{13}$ cm$^{-2}$ for the highest voltages in our measurements. As shown in Fig. 4(a), our calculation reproduces well the phonon absorption amplitude for the two types of trilayers, with strong gate dependence in the ABC system and a weaker dependence in the ABA system. A similar behavior is also observed for the frequency shift of the two modes, as evaluated by the phonon self-energy [Fig. 4(b)]. This similarity reflects the fact that both dependences are related to the virtual interband particle-hole excitations [28, 29]. According to our calculation, the red shift of the phonon frequency arises from coupling of the phonons to electronic transitions of higher energies, the phase space of which increases with doping.

There are numerous possible interband electronic transitions in the trilayers. However, because of the symmetry of the IR active phonons, only a few of them make significant contributions to the IR phonon response. In particular, optical transitions between the low-lying states, such as the v1-c1 transition in Figs. 1(c)-(d), are forbidden by selection rules. Also, the pairs of interband transitions that are symmetric under the electron-hole exchange, such as the v1-v2 and c1-c2, v1-c2 and v2-c1, as well as v1-c3 and v3-c1 transitions, tend to cancel each other in their influence to the IR phonons [28, 29]. As we have adopted a band structure that is symmetric on the electron and hole sides, our theory predicts zero IR activity for the phonons at zero doping and arbitrary temperature. Effective electron-hole transitions will, however, be induced by doping that increases the space of Pauli-allowed transitions and diminishes the destructive interference between different transitions. Our calculation shows that a main contribution to the phonon absorption comes from the electronic transitions between the low- and high-lying conduction (valence) bands, as indicated by the arrows in Figs. 1(c)-(d). The principal distinction between the ABA and ABC electronic structure is thus the energy separation between the low- and high-lying bands, which is approximately $\sqrt{2}\gamma_1 \sim 0.52$ eV and $\gamma_1 \sim 0.37$ eV for ABA and ABC trilayers, respectively. The effect of the coupling with such particle-hole excitations will consequently be considerably stronger in the ABC structure, which is more nearly resonant with the (~0.2 eV) phonon energy.

In our experiment, we observe a weak IR absorption in the ABA trilayer at $V_{CN}$ that is not expected by our theory. Such residual absorption may reflect the inhomogeneous spatial distribution of charges expected in real experimental samples and the asymmetry in the band structure of the ABA trilayer [3, 9] that allows particle-hole excitations to be coupled to the phonons in the absence of doping. Further study of the subject also requires the consideration of the gate induced perpendicular electric field, which is expected to lower the symmetry of the crystals and hence induce IR activity of all three in-plane optical modes even at the charge neutrality point.

In conclusion, our experiment demonstrates that IR response of optical phonons in ABC trilayers is much more sensitive to doping than in ABA trilayers. The observation reflects the different nature of the coupling of the phonons with electronic transitions in the two types of trilayers. The results can be explained by a simple theory that considers only the dominant IR active phonon modes and the influence of band filling effects. Previous studies [21, 22] have also shown an enhanced phonon absorption and energy red shift with doping in the bilayer graphene. Our present results show thus that in this regard the ABC trilayer is rather similar to bilayer graphene, but that the ABA trilayer stands as a new system in its own right. Such contrast in behavior as a function of stacking order parallels that found in studies of gate-induced band-gap opening in bilayer and trilayer graphene [3, 24, 35, 36].

We acknowledge A. B. Kuzmenko for important support in the Kramers-Kronig constrained variational analysis of the optical spectra. E.C. acknowledges the Marie Curie grant PIEF-GA-2009-251904. This work was supported by the National Science Foundation (grant DMR-1106225) and by the graphene MURI program of the Office of Naval Research.

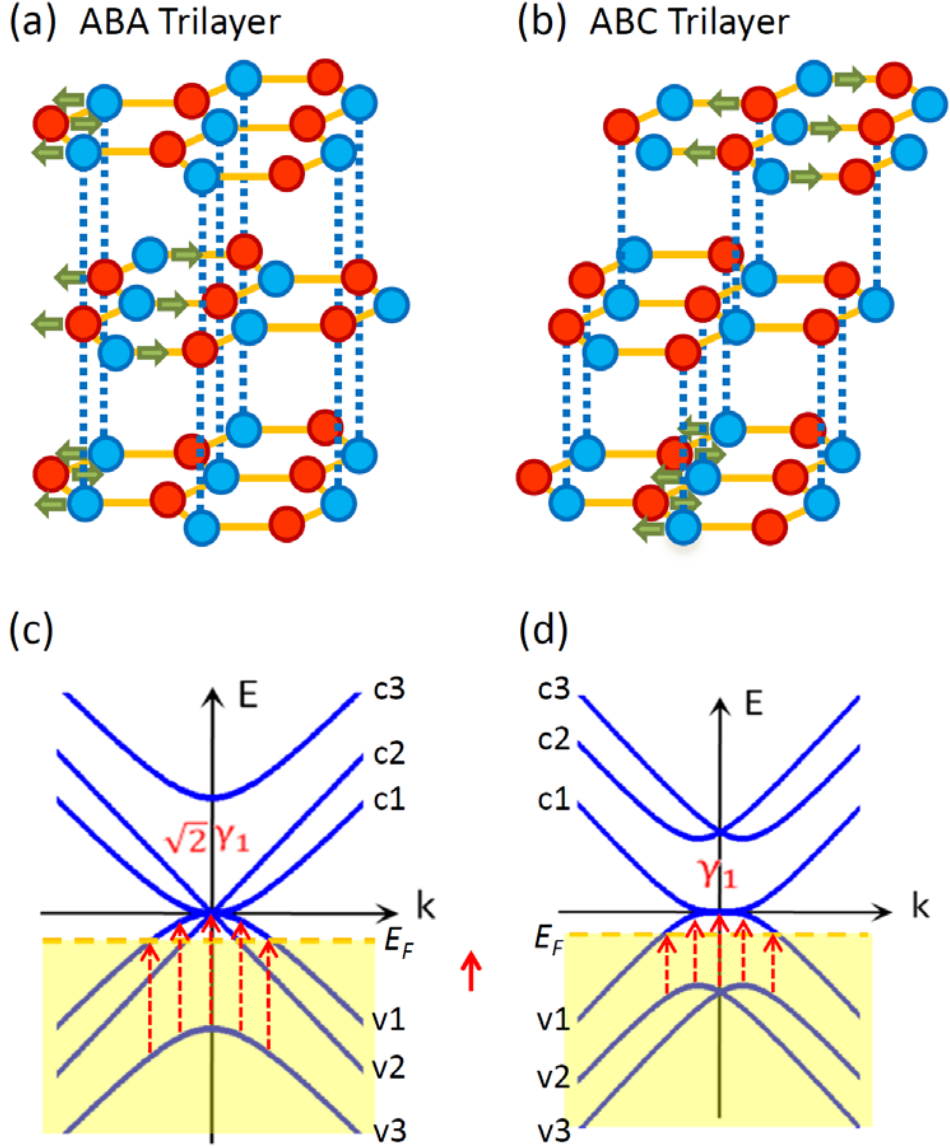

FIG. 1. (a)-(b) Lattice structure of trilayer graphene with ABA (a) and ABC (b) stacking order. The red and blue dots represent carbon atoms in the A and B sublattices of the graphene honeycomb structure. The arrows represent the atomic displacements of the major IR active optical phonon modes: $E'_b$ mode in ABA trilayer and $E_u$ mode in ABC trilayer. (c)-(d) Band structure of graphene trilayers with ABA (c) and ABC (d) structure based on a tight-binding model with only $\gamma_0$ and $\gamma_1$ intra- and inter-layer couplings. The conduction and valence bands are symmetric and the energy gaps between the low-lying band and the high-lying bands are approximately $\sqrt{2}\gamma_1$ and $\gamma_1$ for ABA and ABC trilayer, respectively. The dashed red arrows show the most important virtual transitions that contribute to the renormalization of the IR active phonons at finite doping. The solid red arrow between panels (c) and (d) indicates the energy of the zone-center optical phonon.

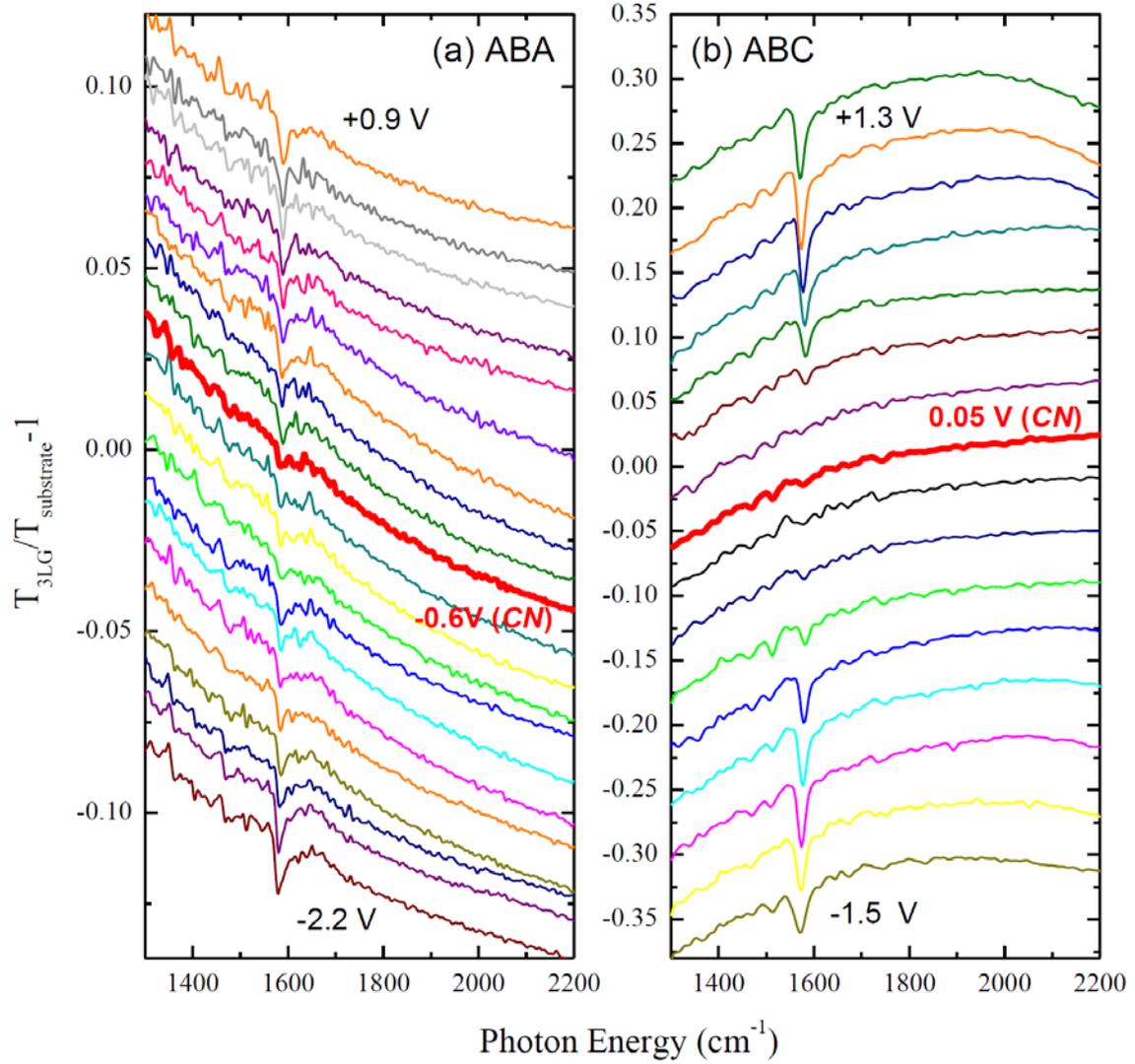

FIG. 2. (a)-(b) Infrared transmission spectra, displaced for clarity, of trilayer graphene with ABA (a) and ABC (b) stacking order for different gate voltages $V_g$. The spectra at the charge neutrality (CN) points, at gate voltage $V_{CN}$, are highlighted. From top to bottom, the gate voltages for the ABA spectra are: $V_g$ = 0.9, 0.7, 0.5, 0.3, 0.1, -0.1, -0.3, -0.4, -0.5, -0.6 (CN), -0.7, -0.8, -0.9, -1.0, -1.1, -1.2, -1.4,-1.6,-1.8,-2.0 and -2.2 V; The gate voltage for the ABC spectra are from top to bottom: $V_g$ =1.3, 1.2, 1.0, 0.8, 0.6, 0.4, 0.2 0.05 (CN), -0.1, -0.3, -0.5, -0.7, -0.9, -1.1, -1.3 and -1.5 V.

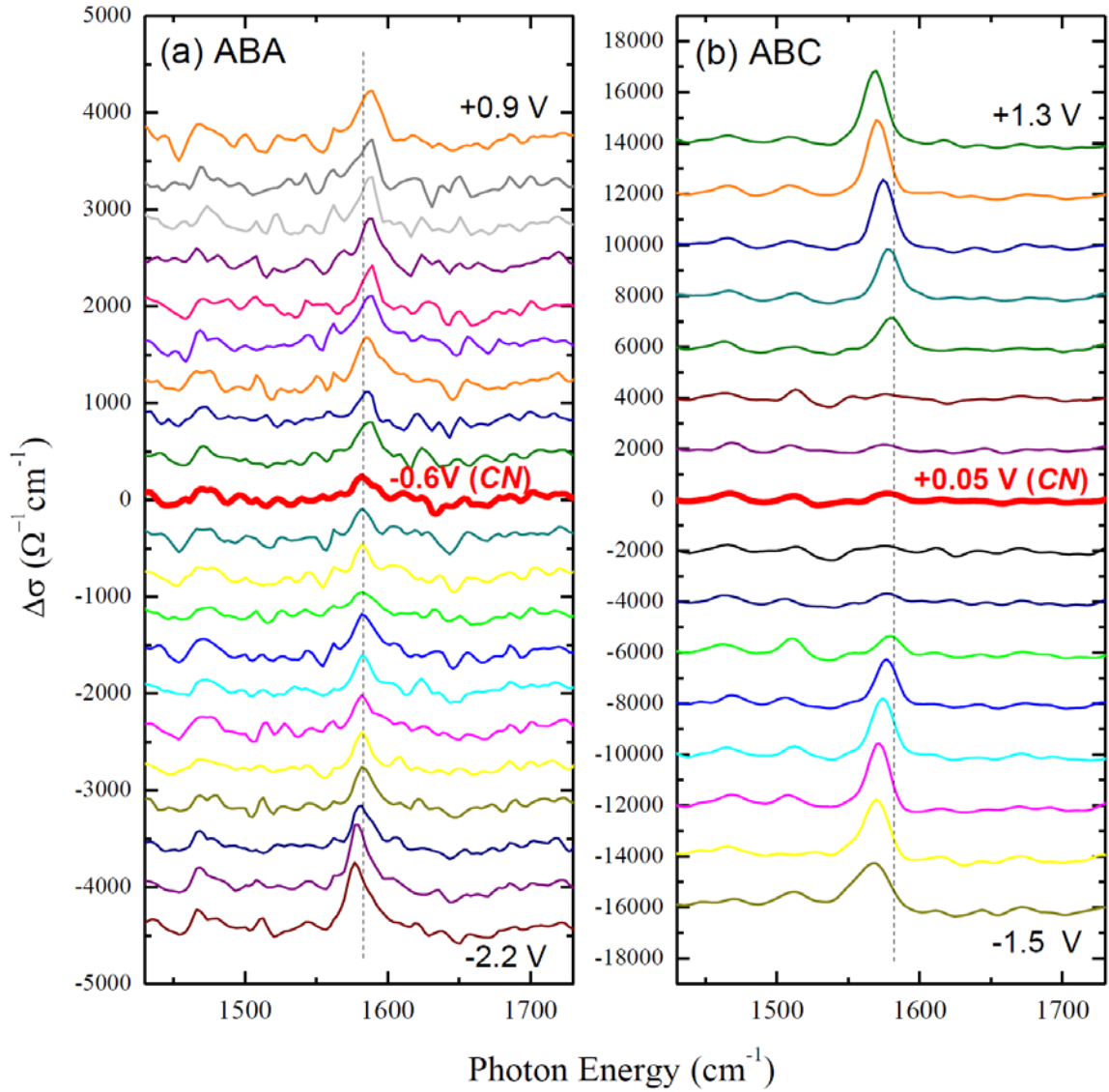

FIG. 3. (a)-(b) The optical conductivity spectra Δσ(ω) extracted from the corresponding baseline-adjusted transmission spectra in Figs. 2(a)-(b). The dashed lines are guides to the eye to show the shift of the phonon energy.

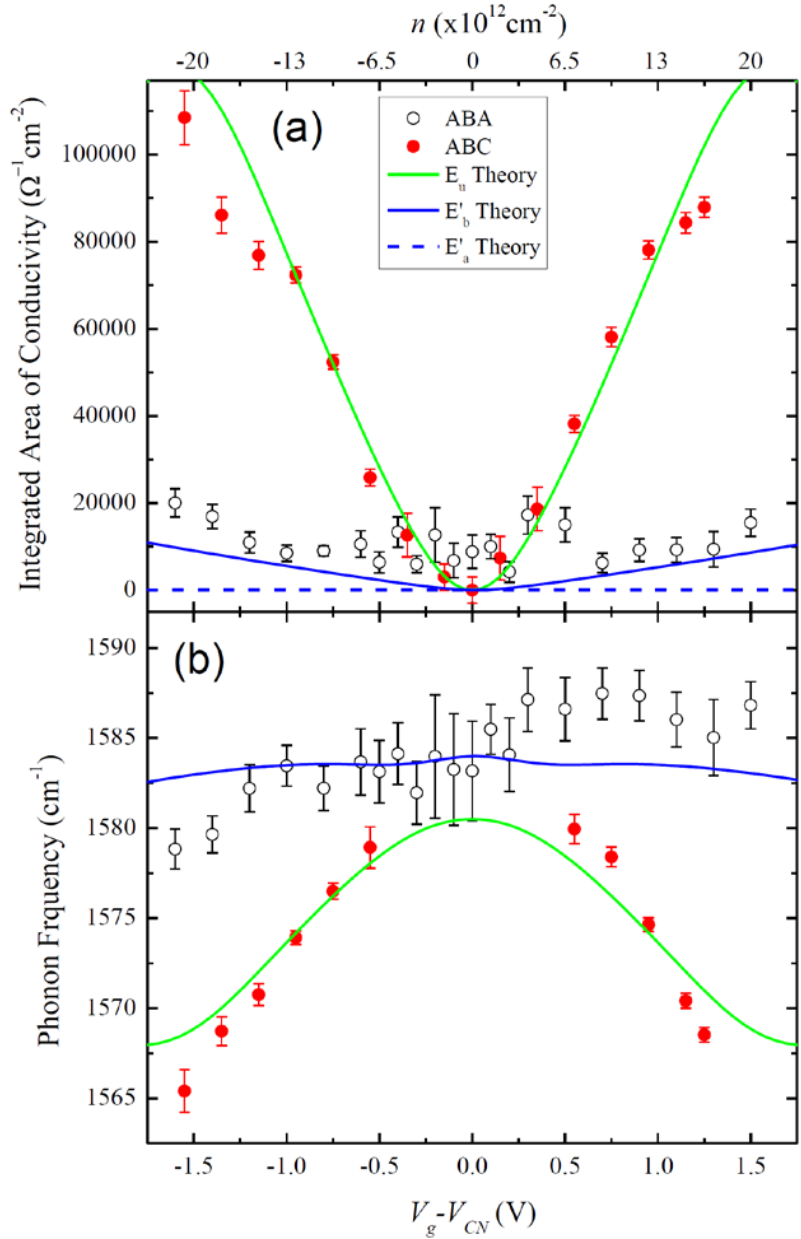

FIG. 4. The integrated optical conductivity (a) and the phonon frequency (b) extracted from the spectra in Figs. 3(a)-(b) by using a single-Lorentzian fitting, displayed as a function of gate voltage ($V_g$, bottom axis) and of the doping charge density ($n$, top axis). The symbols are experimental data and the lines are from the charged-phonon model described in the text. The error bars represent the uncertainty in fitting the spectra. The values of $V_{CN}$ are subtracted from $V_g$ for better comparison of the ABA and ABC data. In (a), we show both the calculated spectral intensity of the $E'_a$ and $E'_b$ modes in the ABA trilayer. We see that the $E'_b$ mode has much higher IR activity than the $E'_a$ mode. For the theoretical fits in (b), we assume zero-doping phonon frequencies of 1584 cm$^{-1}$ and 1580.5 cm$^{-1}$ for ABA and ABC trilayers, respectively.

# Supplemental Material

**Charged-phonon theory for trilayer graphene**

In this supplement material, we briefly describe the implementation of charged-phonon theory, previously applied to bilayer graphene [S1, S2], to the case of trilayer systems for both the ABA and ABC stacking order. In our model, we consider the effect of the gating in terms of a rigid-band doping, disregarding any possible effects of the perpendicular electric field on the band structure itself. This assumption permits us to focus on the few infrared active in-plane optical modes present in the pristine graphene trilayers. The vertical electric field would make all the lattice modes infrared active. As we have shown in the main text, this simplified model still allows us to explain the principal aspects of the gate-dependent phonon spectra for both ABA and ABC stacked trilayers.

To obtain a suitable basis for the lattice eigenvectors, we follow Ref.[S3, S4] by solving the secular equation

$$\begin{pmatrix} 0 & \varepsilon & 0 \\ \varepsilon & \delta & \varepsilon \\ 0 & \varepsilon & 0 \end{pmatrix} \varphi_v = E_v \varphi_v, \qquad (1)$$

with $\varepsilon = 2.2$ cm$^{-1}$ and $\delta = 3$ cm$^{-1}$. Here $E_v$ is the energy eigenvalue (measured with respect to the frequency of the uncoupled in-plane optical mode) and $\varphi_v$ is the eigenvector of the lattice mode $v$. Two lattice modes ($E'_a$ and $E'_b$) are found be infrared active in the ABA trilayers, while only one lattice mode ($E_u$) is infrared active in the ABC structure. The eigenvectors of these lattice modes, denoted by the relative lattice displacements of the carbon atoms vertically connected in each graphene layer, *i.e.* (A1, B2, A3), are:

$$\begin{aligned}
\varphi^T_{E'_a} &= \left[ \varepsilon, \left( \delta - \sqrt{\delta^2 + 8\varepsilon^2} \right)/2, \varepsilon \right], \\
\varphi^T_{E'_b} &= \left[ \varepsilon, \left( \delta + \sqrt{\delta^2 + 8\varepsilon^2} \right)/2, \varepsilon \right], \\
\varphi^T_{E_u} &= [1, 0, -1].
\end{aligned} \qquad (2)$$

The spectral properties of each lattice mode $v$ can be computed in terms of the complex mixed current-phonon response function $\chi_{jv}(\omega)$ [S1, S2]. We calculated the trilayer electronic band structure by employing a standard $\mathbf{k}\cdot\mathbf{p}$ tight-binding model with the nearest-neighbor intralayer coupling $\gamma_0 = 3.16$ eV and interlayer coupling $\gamma_1 = 0.37$ eV. The current operator $j_{k,x}$ along the x-axis is $j_{k,x} = dH/dk_x$ and the electron-phonon interaction Hamiltonian is $H_{el-ph} = g\sum_{\mathbf{k}} \psi_{\mathbf{k}}^{\dagger} V_v \psi_{\mathbf{k}} \phi_v$, which is the linear expansion of the electronic Hamiltonian with respect to the dimensionless lattice displacement operator $\phi_v$ [S5, S6]. The deformation potential has been estimated to be $g = 0.27$ eV [S1, S2, S7, S8]. The peak intensity $W_v$ for each mode is computed as $W_v = \pi\left[\text{Re}\,\chi_{jv}(\omega_v)\right]^2/\omega_v$. In the same way, we can obtain the phonon self-energy as the response function $\chi_{vv}(\omega)$. From the phonon self-energy we compute the phonon frequency shift as $\Delta\omega_v = \text{Re}\,\chi_{vv}(\omega_v)$. All the quantities have been evaluated at a temperature of $T = 300$ K, including a phenomenological damping of $\eta = 20$ meV to account for the disorder and impurities in the sample. (We note that the results do not depend significantly on the value of $\eta$.)

Figure S1 displays our calculated results for the complex current-phonon response function $\chi_{jv}(\omega)$ for the $E'_b$ mode in the ABA trilayer and the $E_u$ mode in the ABC trilayer at different doping densities from $n = 1 \times 10^{12}$ cm$^{-2}$ to $n = 20 \times 10^{12}$ cm$^{-2}$. For all doping levels, we observe a peak of the real part of the response function, which is of similar magnitude for the two types of trilayers. The energy of this peak is, however, different for the two allotropes (~0.37 eV and ~0.52 eV for the ABC and ABC trilayers, respectively), a result that reflects the different band structure of the two materials. The phonon energy lies below these peaks, in a regime where the value of the response function decreases monotonically with decreasing energy. The response at the phonon energy (~0.2 eV) is thus significantly stronger in the ABC trilayer than in the ABA trilayer.

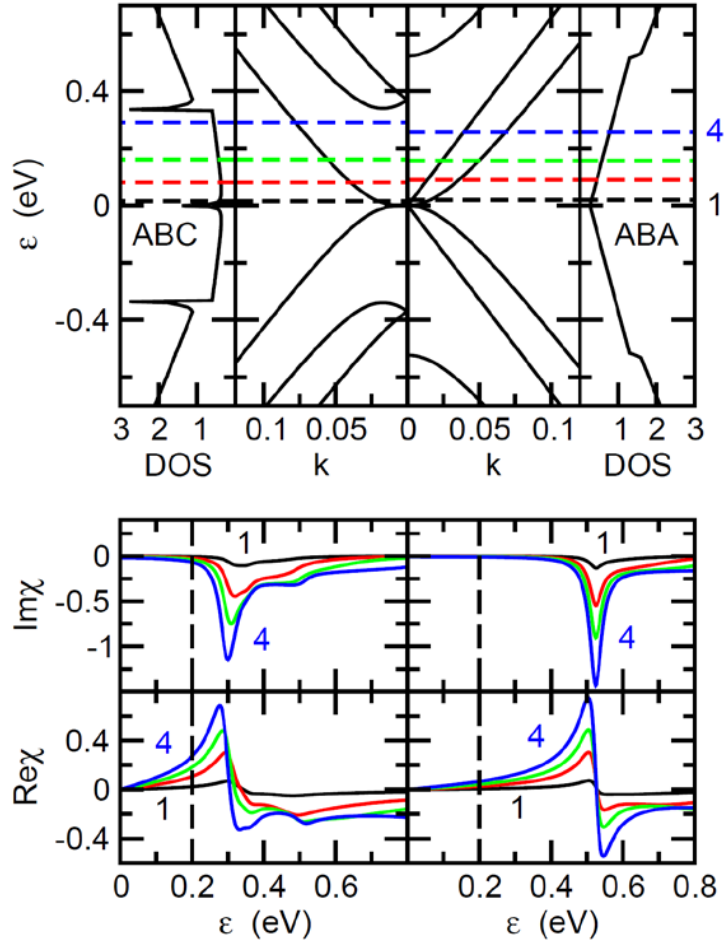

Figure S1. Upper panels: A representation of the electronic band structure and the density of states (DOS) for trilayer graphene with ABC (rhombohedral) and ABA (Bernal) stacking. The dashed lines represent the chemical potentials shifted by gate-induced doping. The doping levels, from bottom to top (1-4), are 1, 5, 10 and 20 x $10^{12}$ cm$^{-2}$. Lower panels: The real and imaginary part of the corresponding mixed current-phonon response function responsible for the IR activity of the optical phonons. For comparison, the dashed lines indicate the energy of the optical phonons.

**Supplemental references:**